\documentclass[twocolumn,prb,superscriptaddress,showpacs]{revtex4}
\newcommand{\tc}{$T_{\rm c}$}
\newcommand{\pcg}{PuCoGa$_5$}
\newcommand{\la}{$\lambda$}

\newcommand{\msr}{$\mu$SR}

\usepackage{graphicx}
\usepackage{dcolumn}
\usepackage{bm}

\begin{document}


\title{Muon spin rotation measurements of the superfluid density in fresh and aged superconducting PuCoGa$_5$
}

\author{K.~Ohishi}
\affiliation{Japan Atomic Energy Agency, Tokai, Ibaraki 319-1195,
Japan}
\author{R.H.~Heffner}
\affiliation{Japan Atomic Energy Agency, Tokai, Ibaraki 319-1195, 
Japan}
\affiliation{Los Alamos National Laboratory, Los Alamos, New
Mexico 87545 USA}
\author{G.D.~Morris}
\affiliation{Los Alamos National Laboratory, Los Alamos, New
Mexico 87545 USA}
\affiliation{TRIUMF, 4004 Wesbrook Mall, Vancouver, B.C., Canada V6T 2A3}
\author{E.D.~Bauer}
\affiliation{Los Alamos National Laboratory, Los Alamos, New
Mexico 87545 USA}
\author{M.J.~Graf}
\affiliation{Los Alamos National Laboratory, Los Alamos, New
Mexico 87545 USA}
\author{J.-X.~Zhu}
\affiliation{Los Alamos National Laboratory, Los Alamos, New
Mexico 87545 USA}
\author{L.A.~Morales}
\affiliation{Los Alamos National Laboratory, Los Alamos, New
Mexico 87545 USA}
\author{J.L.~Sarrao}
\affiliation{Los Alamos National Laboratory, Los Alamos, New
Mexico 87545 USA}
\author{M.J.~Fluss}
\affiliation{Lawrence Livermore National Laboratory, P.O. Box 808,
Livermore, California 94550 USA}
\author{D.E.~MacLaughlin}
\affiliation{Department of Physics, University of California, Riverside,
California 92521 USA}
\author{L.~Shu}
\affiliation{Department of Physics, University of California, Riverside,
California 92521 USA}
\author{W. Higemoto}
\affiliation{Japan Atomic Energy Agency, Tokai, Ibaraki 319-1195, 
Japan}
\author{T.U.~Ito}
\affiliation{Japan Atomic Energy Agency, Tokai, Ibaraki 319-1195,
Japan}
\affiliation{Department of Physics, Tokyo Institute of Technology,
Megro-ku, Tokyo 152-8551, Japan}

\date{\today}

\begin{abstract}
We have measured the temperature dependence and magnitude of the
 superfluid density $\rho_{\rm s}(T)$ via the magnetic field 
penetration depth $\lambda(T)$ in PuCoGa$_5$ (nominal critical 
temperature $T_{c0} = 18.5$ K) using the muon spin rotation 
technique in order to investigate the symmetry of the order parameter, 
and to study the effects of aging on the superconducting properties 
of a radioactive material. The same single crystals were measured 
after 25 days ($T_c = 18.25$ K) and 400 days ($T_c = 15.0$ K) of 
aging at room temperature. Penetration depths 
$\lambda(0) =$~ 265(5) and $\ge$ 498(10) nm are derived for the fresh and 
aged samples, respectively. The temperature dependence of the 
superfluid density $\rho_{\rm s}(T)/\rho_{\rm s}(0)=\lambda(0)^2/\lambda(T)^2$ 
is well described in both materials by a model using 
$d$-wave gap symmetry. Within the context of a strong-coupling, dirty $d$-wave model, 
a zero-temperature gap value 
$\Delta_0 = 3.0k_BT_{c0}$ is obtained in the fresh sample for a scattering rate $\Gamma=0.005 \pi k_BT_{c0}$, which is consistent with Abrikosov-Gor'kov (AG) pair-breaking theory. This $\Delta_0$ should be compared to 
the weak-coupling value for a clean $d$-wave gap $\Delta_0 = 2.14k_BT_{c0}$.
In the aged sample the same model yields $\Delta_0 = 2.4k_BT_{c0}$ for $\Gamma=0.010 \pi k_BT_{c0}$. This value of $\Gamma$ is much less than required by the AG pair-breaking formalism. Furthermore, the aged $\rho_{\rm s}(0)$ is reduced by at least 70\% compared
to the fresh sample, which is also incompatible with $\Delta T_c/T_{c0} \sim$ 20\%, according to 
AG theory. We conclude that the data in aged PuCoGa$_5$ support the
postulate that the scattering from radiation-induced defects is not in
the limit of the AG theory of an order parameter which is spatially averaged over impurity sites, but rather in the limit of short-coherence-length superconductivity. We show that a model calculation consistent with this assumption
fits $\rho_{\rm s}(T)$ in the aged sample rather well.

\end{abstract}

\maketitle

\section{Introduction}
Interest in the magnetism and superconductivity of $f$-electron
materials has remained strong over the last several decades, starting
with the discovery\cite{HF} of heavy fermion superconductivity in Ce-
and U-based materials, and continuing today with the
investigation\cite{PetrovicYEAR01} of novel ground states in the
tetragonal Ce-based CeTIn$_5$ (T=Co, Ir, Rh) materials. Pu-based
compounds exhibit a particular richness and complexity, as is
illustrated by the discovery \cite{SarraoYEAR02} of superconductivity in
\pcg\/ at an order of magnitude higher transition temperature ($T_{c0} =
18.5$ K) than the structurally similar heavy fermion 
superconductor\cite{PetrovicYEAR01} CeCoIn$_5$ (\tc\/$\sim$2.3~K). One
reason for this complexity in Pu materials is that their $f$ electrons
sit at the boundary between more fully localized (as in Am compounds) and more fully itinerant
(as in U and Np compounds) behavior.\cite{Pu} 

There is evidence that the superconductivity in PuCoGa$_5$ may be
unconventional in nature. Recent nuclear magnetic resonance (NMR) and
nuclear quadrupole resonance (NQR) studies\cite{CurroYEAR05} display an
absence of a coherence peak, together with a decreasing Knight shift and
a $T^3$-behavior in the spin-lattice relaxation rate $1/T_1$ just below
$T_c$, suggesting that the superconducting order parameter has $d$-wave
symmetry, as in the copper-oxide
superconductors.\cite{KirtleyYEAR06}
Calculations\cite{TanakaYEAR04,HottaYEAR03,OpahleYEAR03} of the
electronic structure of \pcg\/ show a Fermi surface consisting of
several cylindrical sheets, which is favorable for $d$-wave,
spin-fluctuation mediated superconductivity. It has been
suggested\cite{CurroYEAR05} that \pcg\/ belongs to a class of
spin-fluctuation induced superconductors, lying between the
high-temperature, copper-oxide superconductors and the low-temperature,
heavy-fermion superconductors on a plot of transition temperature
versus spin fluctuation temperature.\cite{CurroYEAR05,MoriyaYEAR03} 

Additional measurements which test this supposition are, therefore,
important. This is particularly true because the natural radioactivity
of Pu ($^{239}$Pu half-life = $24,000$ years) creates lattice defects
which scatter electrons and can create impurity bands which partly
obscure the signature of a pure $d$-wave superconductor. Such effects
are evident in the low-temperature NQR spin-lattice relaxation rate
$1/T_1$,\cite{CurroYEAR05} where the $T^3$ temperature dependence gives
way to a linear-$T$ behavior below about $0.4 T_c$. As we show, the
temperature dependence of the
penetration depth $\lambda(T)$ in PuCoGa$_5$ is less sensitive to these defects, thus
providing a clean test of the gap symmetry and of the robustness of the
order parameter to the presence of pair-breaking defects. 
The magnitude of $\lambda(0)$ 
is strongly sensitive to defects, however, a fact which we are able to explain in terms of the 
relatively short superconducting coherence length in PuCoGa$_5$.
Finally, the study of
radiation effects in superconductors is of practical importance as well,
because increases in the critical current density can be achieved by
the pinning of flux lines at defects.

In this paper we present transverse-field (TF) muon spin rotation
($\mu$SR) measurements of the in-plane magnetic field penetration depth
$\lambda$ in the same single crystals of \pcg\/ after 25 days and
400 days of aging at room temperature. $\lambda(T)$ is determined by the spectrum of
quasiparticle excitations exceeding the superconducting energy gap
$\Delta(T)$, and is thus a sensitive measure of the temperature
dependence of the superfluid density $\rho_{\rm s}(T) \propto
\lambda(T)^{-2}$, and the gap structure. Some of the data after
25 days of aging has been reported previously. \cite{MorrisYEAR06} 
 
\section{Experimental details}
\subsection{Sample preparation and experimental setup}
Two crystals of PuCoGa$_5$ measuring $\sim 5 \times 6$ mm$^2$ and about $1/2$ mm thick
were grown from excess Ga flux \cite{SarraoYEAR02} as flat plates with
the $c$-axis normal to the surface. The crystals were encapsulated in a
polyimide coating about 70 $\mu$m thick to prevent contamination from
particle and ejecta emission. The encapsulated crystals were then
attached and sealed under a helium atmosphere inside a Ti cell for further
protection. The cell was cooled using a continuous-flow He cold-finger
cryostat. 

The experiments were performed at the M20 surface muon channel at
TRIUMF, Vancouver, Canada. Muons entered the cell through a 50 $\mu$m
Ti-foil window, with their polarization rotated vertically 90$^\circ$,
perpendicular to the beam momentum. The applied field $H_0$ was along
the incoming beam direction in the TF mode, and parallel to the
crystalline $c$-axis. A negligible fraction of the beam stopped in the
Ti window and polyimide coating.

The initial set of measurements (after 25 days) were performed with the
sample mounted directly on the Ti-cell backing with $H_0 = 60$ mT. 
The second set of measurements (after 400 days, the exact time determined by TRIUMF's beam-time schedule) was carried out in the
same geometrical and spin configurations as before, but with the sample
surrounded by a Ag backing plate and in applied fields of 60 mT and
0.3 T. All data with $T \leq T_c$ were taken in a field-cooled
mode.

\subsection{Data analysis}
In an ideal experiment the muons stop randomly on the scale of the flux
line lattice (FLL) spacing. Therefore, the muon spin precession signal
$\hat{P}(t)$ provides a random sampling of the internal field distribution $n(B)$,\cite{BrandtYEAR1988}
\begin{eqnarray}
\hat{P}(t) \equiv P_x(t)+iP_y(t)=\int_{-\infty}^\infty
n(B)\exp(i\gamma_\mu Bt)dB,
\end{eqnarray}
where $n(B)=\langle\delta(B-B(r))\rangle_r$ is the spatial average of a Dirac 
delta function, $\gamma_\mu$ (=
$2\pi\times$135.53~MHz/T) is the muon gyromagnetic ratio and $B(r)$ is
the internal field. This equation indicates that the real amplitude of
the Fourier transformed muon precession signal corresponds to the
spectral density $n(B)$. 
In general, an asymmetric $n(B)$, the so-called `Redfield pattern', is
expected when the FLL consists of straight, rigid flux lines and the
ratio $\kappa = \lambda/\xi$ is not too large.\cite{BrandtYEAR88} In
such a case, $n(B)$ possesses a high-field tail which characterizes the
flux-line core radius (or coherence length $\xi$); the second moment
$\langle (\Delta B)^2\rangle$ of $n(B)$ determines the London
penetration depth.
 
\begin{figure}[htbp]
\rotatebox[origin=c]{0}{\includegraphics[width=0.85\columnwidth]{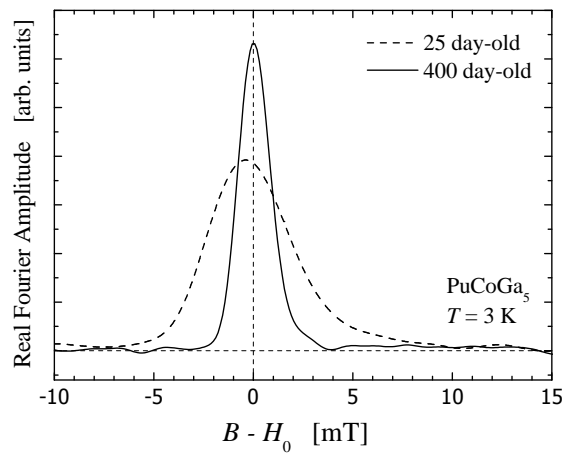}}
\caption{The fast Fourier transform spectra of the \msr\/ signals in
 \pcg\/ at 3~K under 60~mT, after subtracting the
 sample holder signal, in both the 400 day-old sample and the 25 day-old
 sample. $H_0$ was determined from a fit to the normal-state data.}
\label{FFT}
\end{figure}
The fast Fourier transform (FFT) provides a reasonably good
representation of $n(B)$ in the FLL state. Fig.~\ref{FFT} shows FFT
spectra for the \msr\/ signals from the sample after 25 and 400 days at
$T = 3$ K. Here the background signal from the sample cell has been
subtracted from the time spectrum before carrying out the FFT. Some
age-related differences in the spectra are apparent. First, while the
fresh sample exhibits a slightly asymmetric line shape and a small
negative field shift, the aged sample shows a nearly symmetric shape
with almost no negative shift. Second, the linewidth has narrowed
appreciably with aging. 

Attempts to fit the fresh sample data using standard models
\cite{SonierYEAR00,KadonoYEAR04} which incorporate a broadened (due to
instrumental and nuclear line-width effects) Redfield pattern for a
perfect FLL were not successful. This is due to the fact that the
lineshape is only moderately asymmetric, and thus unique determinations
of $\lambda$ and the core radius, which depend upon having a high-field
tail in the field distribution, cannot be obtained. Most of the
decreased asymmetry in the lineshape is due to the fact that PuCoGa$_5$
is a high $\kappa~ (>100)$ superconductor with a relatively large penetration depth (see below), but part may also be due to
distortions of the FLL from radiation induced pinning centers, as we
discuss below.

As described earlier, reasonable fits to the fresh-sample data were
obtained with simple Gaussian or Lorentzian line shapes, and these two
functional forms yielded the same temperature dependence for the
linewidths.\cite{MorrisYEAR06} Nevertheless, in an attempt to account
phenomenologically for any small asymmetry, we also performed fits using
a sum of two Gaussians with different centroid frequencies and
widths.\cite{KhasanovYEAR05} When these widths were convoluted into a single
width (as in Ref. \onlinecite{KhasanovYEAR05}) the same temperature dependence and
overall linewidth was obtained as for the single Gaussian fits, but with
somewhat larger ($\cong 10-15$\%) uncertainties. 

\begin{figure}
\centerline{\includegraphics[width=.85\columnwidth]{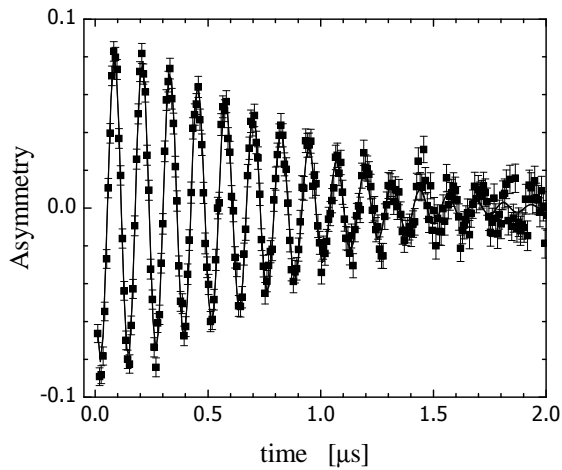}}
\caption{
Muon TF precession signal at $T = 4$~K, corresponding to the first term in Eq. (\ref{asym}). The background component was characterized at times between $t = 3 - 10$ $\mu$s, where the sample signal is fully relaxed, and
subtracted to show the inhomogeneous relaxation produced by
 the flux lattice in the superconducting state. The solid line is a fit
 for a Gaussian $P(t)$ in Eq. (\ref{asym}).
}
\label{Gauss-osc}
\end{figure}

Accordingly, the TF precession spectra for both samples were fit to the
sum of two terms, corresponding to muons stopping in the sample and
background materials (either Ti or Ag), respectively: 
\begin{eqnarray}
A_0 G_{\rm z}(t) &=& A \cos(\omega t + \phi) P(t) \nonumber \\
&+& A_{\rm b} \cos(\omega_{\rm b}t + \phi)\exp(-\sigma_{\rm b}^2 t^2/2),
\label{asym}
\end{eqnarray}
where $A$ and $A_{\rm b}$, $\omega$ and $\omega_{\rm b}$ are the partial
asymmetries and central frequencies for \pcg\/ and the sample holder,
respectively, $A_0$ is the total positron decay asymmetry 
($A_0=A + A_{\rm b}$), and $\phi$ is the initial phase.\cite{Note1} Here we used 
$P(t) = \exp(-\sigma^2t^2/2)$. Fits to the data yielded $A/A_{\rm b} \cong 1/2$. 
The background signal from Ti in $H_0$ = 60 mT applied field was well
characterized by a Gaussian relaxation function,
$G_b(t)=\exp(-\sigma^2_bt^2/2)$, with $\sigma_b \approx$ 0.014
$\mu$s$^{-1}$. The relaxation rate from the Ag backing used in the
measurements on the aged sample is negligible.

The quality of the Gaussian fits for the 25 day-old sample is
illustrated in Fig.~\ref{Gauss-osc} for data taken at $T = 4$ K. 
Here the $G_{\rm z}(t)$ data (Eq. (\ref{asym})) from $t = 3-10$ $\mu$s
were fit separately, and this long-time Ti signal was then subtracted
from the total spectrum, leaving only the signal from the sample in the
superconducting state. One sees that the Gaussian form for $P(t)$ gives
a satisfactory fit. Thus, we conclude that single Gaussian fits give an
acceptably accurate measure of the linewidths in the 25 day-old sample.
Regarding the 400 day-old data, a single Gaussian fit is obviously
appropriate, as can be seen in Fig.~ \ref{FFT}. 

\section{Results}
\subsection{Flux Pinning}
\begin{figure}[htbp]
\rotatebox[origin=c]{0}{\includegraphics[width=.75\columnwidth]{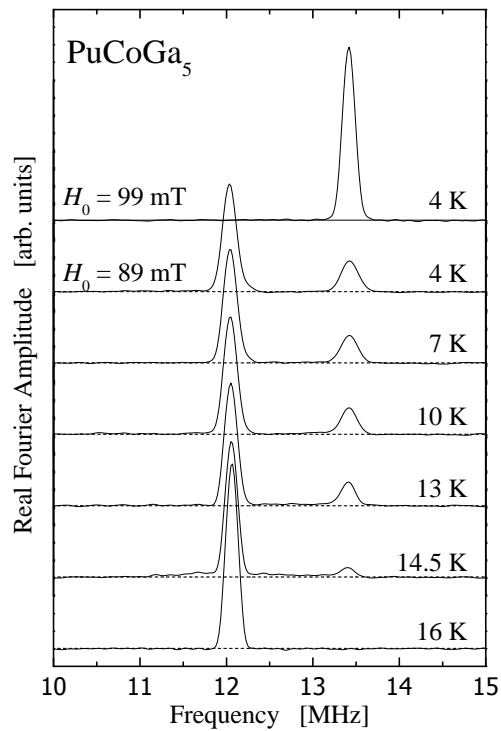}}
\caption{Temperature dependence of the real Fourier amplitude [transform of $A_0G_z(t)$] in the 400
 day-old sample showing flux trapping. In the top figure the sample was
 field cooled in $H_0 = 99$ mT. In the lower figures the field has
 been reduced to 89 mT with trapped flux giving rise to a resonance
 at $\approx 13.4$ MHz. 
}

\label{pin}
\end{figure}

We also performed a series of measurements in the aged sample to observe
the effects of flux pinning. The sample was cooled to 4 K in $H_0=99$ mT 
field and a spectrum was obtained. The field was then reduced
by 10 mT without changing the temperature and another spectrum was
taken. The sample was then warmed in $H_0 =89$ mT field, and several
spectra were accumulated at increasing temperatures. The results are
shown in Fig.~ \ref{pin}. In the field-shifted spectrum at 4 K one
observes an unshifted line with an amplitude of 90-100\% of the sample
signal, together with a line at reduced frequency 
from muons stopping in the sample backing material. This clearly
demonstrates strong pinning, and is consistent with the
temperature-independent, field-cooled susceptibilities found in
fresh\cite{SarraoYEAR02} and aged\cite{JutierYEAR2005} samples of
PuCoGa$_5$. As seen in Fig.~ \ref{pin}, the trapped flux gradually leaks
out of the sample as the temperature is raised. Note, however, that for
the data in Fig.~ \ref{sgmvsT}, where the field is held constant at the
field-cooled value, the sample asymmetry $A$ is independent of
temperature, as expected.

\subsection{Temperature dependence of $\sigma$}
\begin{figure}[htbp]
\rotatebox[origin=c]{0}{\includegraphics[width=.75\columnwidth]{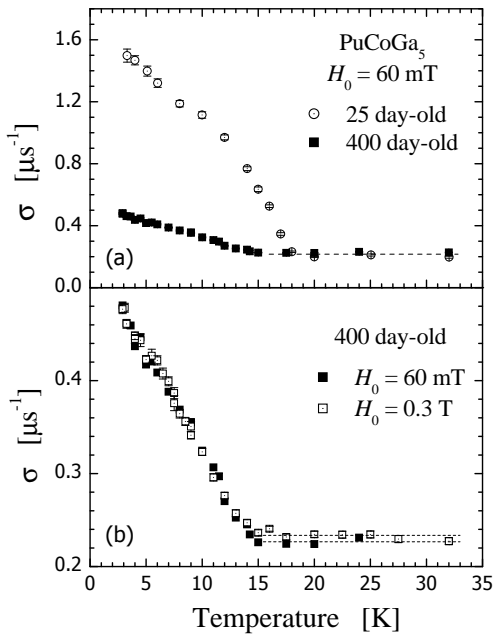}}
\caption{(a) Temperature dependence of Gaussian muon spin relaxation
 rate $\sigma$ for $H_0 = 60$ mT in the 25 day-old and 400 day-old
 samples. The dotted line is a guide to the eye. (b) Temperature
 dependence of $\sigma$ in the 400 day-old sample for $H_0 = 60$ mT and
 0.30 T.
}
\label{sgmvsT}
\end{figure}
The temperature dependences of $\sigma$ for the fresh and aged samples
for $H_0 = 60$ mT are shown in Fig.~\ref{sgmvsT}(a). In both samples
$\sigma$ increases with decreasing temperature below $T_c$ due to the
formation of the FLL. This increase occurs below about 18 K in the fresh
sample, in good agreement with the measured $T_c$. In the 400 day-old
sample the data are consistent with a decrease in $T_c$ of about 3 K, in
reasonable agreement with the radiation-induced reduction of $T_c$ ($\approx 
0.24$ K/month) reported for PuCoGa$_5$ samples of slightly different
isotopic concentrations.\cite{JutierYEAR05}

As seen in Fig.~\ref{sgmvsT}(a), there is no temperature dependence to
the normal state values of $\sigma$ (denoted $\sigma_{\rm n}$
below). Assuming the muons occupy the same sites\cite{SchenckYEAR02} as
in CeRhIn$_5$ we estimate a nuclear dipolar linewidth of about 0.21
$\mu$s$^{-1}$ for both of the suggested
$(\frac{1}{2},\frac{1}{2},\frac{1}{2})$ and $(0,\frac{1}{2},0)$ sites,
which is close to the average of the measured values $\sigma_{\rm n}
\approx 0.2$ $\mu$s$^{-1}$. Furthermore, there is little change in 
$\sigma_{\rm n}$ with aging (Fig.~\ref{sgmvsT}(a)), indicating that
additional aging has little effect on the muon site(s). None of our
conclusions regarding the superconducting properties of PuCoGa$_5$
depend on knowing these exact sites, however. 

We also performed measurements with $H_0 = 0.3$ T in the 400 day-old
sample to investigate the field dependence of $\sigma$. In some
superconductors $\sigma$ displays a non-monotonic field dependence which
has been attributed\cite{NiedermayerYEAR02} to motion of the FLL in low
fields where the interactions between the flux lines is weak due to
their large separation. Figure~\ref{sgmvsT}(b) shows that both the
magnitude and temperature dependence of $\sigma$ remain essentially
unchanged for 60 mT $\leq H_0 \leq 0.3$ T in the 400 day-old sample. Note
that $H_0 = 60$ mT is at least twice the lower critical field $H_{\rm
c1}= \Phi_0\ln(\lambda/\xi)/4\pi\lambda^2 \approx 0.035$ T reported
previously \cite{SarraoYEAR02} and $4-5$ times that implied by our
measurements of $\lambda(0)$ in the fresh sample, as discussed
below. Here $\Phi_0 = hc/2e = 2.07 \cdot 10^{-15}$ Tm$^2$ is the
magnetic flux quantum.

\subsection{Temperature dependence of $\rho_{\rm s}$ and $\lambda$}
The $\mu$SR linewidth from the vortex lattice $\sigma_{\rm v}$ is
obtained by subtracting the temperature-averaged normal-state linewidth
$\overline{\sigma}_{\rm n}$ from the total linewidth $\sigma$ in quadrature:
$\sigma_{\rm v}^2 = \sigma^2 - \overline{\sigma}_{\rm n}^2$. The
temperature-averaged $\overline{\sigma}_{\rm n}$ = 0.204(3) $\mu$s$^{-1}$ in
the fresh sample at $H_0 = 60$ mT, and 0.227(3) $\mu$s$^{-1}$ and
0.234(3) $\mu$s$^{-1}$ for $H_0 = 60$ mT and 0.3 T, respectively, in
the aged sample.

\begin{figure}[htbp]
\rotatebox[origin=c]{0}{\includegraphics[width=0.75\columnwidth]{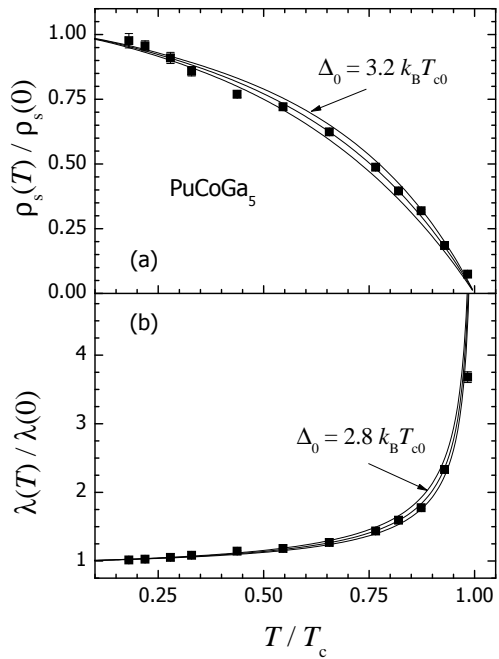}}
\caption{Temperature dependence of (a) the normalized $\mu$SR linewidth
 $\sigma_{\rm v}(T) \propto \rho_s(T)$ and (b) penetration depth
 $\lambda(T)$ for $H_0 = 60$ mT in PuCoGa$_5$ after 25 days aging. 
The solid lines are fits using the $d$-wave model in the strong
 scattering limit described in the text for $\Delta_0 = 2.8, 3.0,$ and
 3.2 $k_BT_{c0}$, where $T_{c0} = 18.5$ K, $T_c = 18.25$ K, and
 scattering rate $\Gamma = 0.005 \pi k_BT_{c0}$. 
$\sigma_{\rm v}(0) = 1.52(5)$ $\mu$s$^{-1}$ and $\lambda(0) = 265(5)$ nm. }
\label{fresh}
\end{figure}

\begin{figure}[htbp]
\rotatebox[origin=c]{0}{\includegraphics[width=0.75\columnwidth]{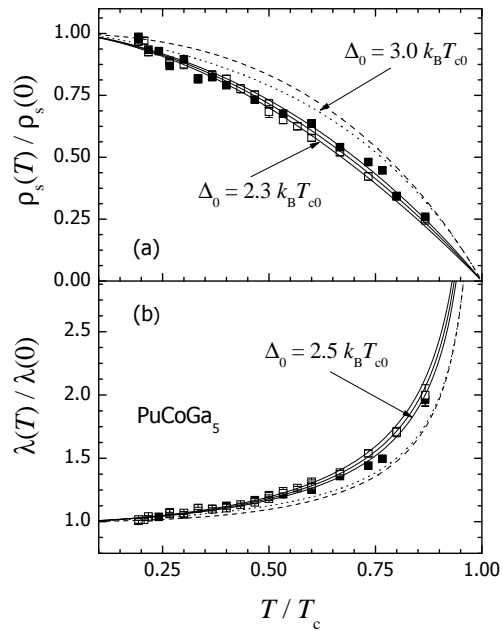}}
\caption{Temperature dependence of (a) the normalized $\mu$SR linewidth
 $\sigma_{\rm v}(T) \propto \rho_s(T)$ and (b) penetration depth
 $\lambda(T)$ in PuCoGa$_5$ after 400 days aging. The solid squares are
 for $H_0 = 60$ mT and the open squares for $H_0 = 0.30$ T. The solid
 lines are fits using the $d$-wave model in the strong scattering
 limit described in the text for $\Delta_0 = 2.3, 2.4,$ and 2.5
 $k_BT_{c0}$, where $T_{c0}=18.5$ K, $T_c = 15.0$ K and fitted scattering rate
 $\Gamma = 0.01 \pi k_BT_{c0}$. $\sigma_{\rm v}(0) = 0.43(1)$ $\mu$s$^{-1}$
 and $\lambda(0) = 498(10)$ nm. The dotted curve is the best fit to the
 fresh sample data in Fig.~ \ref{fresh}. The dashed curve is for
 $\Delta_0 = 2.4k_BT_{c0}$ and $\Gamma=0.074 \pi k_BT_{c0}$. }
\label{aged}
\end{figure}

The penetration depth $\lambda(T)$ can be deduced from $\sigma_{\rm
v}(T)$ as follows:\cite{BrandtYEAR88, KadonoYEAR04} 
\begin{equation}
\sigma_{\rm v}(\mu {\rm s}^{-1})= 48300(1-h)[1+3.9 (1-h)^2]^{1/2}{\lambda^{-2}({\rm nm})}, 
\label{lambda}
\end{equation}
where $h = H_0/H_{\rm c2}(0)$ and $H_{\rm c2}(0) \cong 74$ T is the
upper critical field in the fresh\cite{SarraoYEAR02} sample, so $h \ll
1$. ($H_{\rm c2}$ is larger in the aged sample.\cite{JutierYEAR05})
Eq. (\ref{lambda}) is valid for an isotropic extreme type-II
superconductor (\la$\gg\xi$) with a hexagonal FLL. The $\mu$SR rate
$\sigma_{\rm v}$ measures the rms width $\langle\Delta B^2\rangle$ of
the field distribution from the FLL: $\sigma_{\rm
v}=\gamma_{\mu}\langle\Delta B^2\rangle^{1/2} \propto \rho_{\rm s}/m^*$,
where $m^*$ is the in-plane effective mass. In the following we assume
that the temperature dependence of $\sigma_{\rm v}$ reflects the
temperature dependence of $\rho_{\rm s}$, i.e., $m^*$ is
temperature-independent. Figs. \ref{fresh} and \ref{aged} display the
temperature dependence of $\rho_{\rm s}(T)$ and $\lambda(T)$ for the
fresh and aged samples, respectively, normalized to their values at $T =
0$. The choice of normalization is described below. 

\subsection{Possible effects of flux-line lattice distortion}
As stated above, the $\mu$SR linewidth is directly related to the penetration depth assuming a perfect FLL. In 
real materials, however, some distortion of the lattice always occurs due to slight misalignments and pinning of the flux lines at defect centers. Therefore, before discussing our data in terms of possible theoretical models, it is important to address these issues in PuCoGa$_5$, particularly because radiation damage produces pinning centers. 

It is generally agreed that if the FLL is stable, that is, does not melt or move with temperature, then the temperature dependence of $\sigma_{\rm v}(T)$ accurately reflects the temperature dependence of $\rho_s(T)$. The small scatter in the data observed in Fig.~\ref{sgmvsT} is fully compatible with statistical fluctuations (where Gaussian statistics allow a $\sim$ 37\% probability for fluctuations greater than one standard deviation from the mean), and with non-statistical errors associated with data fitting, and thus is not evidence for FLL motion. Furthermore, as discussed above, both susceptibility and $\mu$SR measurements are consistent with strong pinning in PuCoGa$_5$. These facts, and the absence of a 
field dependence for $\sigma(T)$ in the aged sample, lead to the conclusion that the temperature dependence of $\sigma_{\rm v}(T)$ provides a good measurement of $\rho_s(T)/\rho_s(0)$. 

The second issue concerns the absolute magnitude of $\sigma_{\rm v}(T)$. It is evident from Fig. \ref{sgmvsT} that there is a strong depression of $\sigma_{\rm v}$ in the aged sample ($\approx$ 70\%) despite only a modest 15-20 \% reduction of $T_c$. Conventional pair-breaking theory, originally developed by Abrikosov and Gor'kov (AG),\cite{AGYEAR61} would predict
about a 40\% reduction in $\rho_{\rm s}$ for a reduction in $T_c$ of
20\%.\cite{KimYEAR94,FranzYEAR97} It is, therefore, important to
investigate other possible causes for the reduced linewidth besides
impurity scattering. These include: 1) significant normal-state
inclusions in the aged sample which might give rise to linewidth
narrowing and 2) the effects of distortions to the FLL caused by pinning
at defects.

We first discuss the small probability of large scale normal-state
inclusions, noting that a definitive answer to this question requires a
detailed knowledge of the number and morphology of the sample defects
following room-temperature aging, which is not known with certainty. A $^{239}$Pu
nucleus decays into a 5 MeV alpha particle and an 86 keV recoiling $^{235}$U atom. Most of
the damage cascade is caused by the recoiling heavy U atom which has a
range of $\approx 17$ nm.\cite{JutierYEAR2005} 
Lattice vibrations at room
temperature usually cause some of this damage to be
self-annealed.\cite{McCallYEAR05} Note, however, that vacancy clusters or
dislocation loops with radii $\leq \xi \sim$ 2 nm retain their
superconducting character via the proximity effect.\cite{Tinkham} Unfortunately, reliable
TEM studies of aged PuCoGa$_5$, which could visualize the damage, are
presently lacking. We note, however, that the field-shifting
experiments shown in Fig.~\ref{pin} do not show any evidence for flux
trapped inside normal-state inclusions, because the asymmetry of the
unshifted line is about equal to that measured in a field-cooled
state. Given the sensitivity of our measurements, we thus estimate an
upper limit for the normal-state fraction in our aged sample of $\leq
10$ \%, which is slightly less than, but of the same order as, estimated
from recent EXAFS experiments.\cite{Booth2007} Hence, we find little compelling evidence
for large-scale, normal-state inclusions in the aged sample, and dismiss
this effect as a significant cause of the observed linewidth narrowing.

We now discuss possible effects of FLL distortion by pinning of
vortices. Normally one expects that disorder in the FLL will lead to
linewidth {\em broadening}. This is in fact predicted for the random
pinning of stiff, relatively straight vortex lines.\cite{BrandtYEAR88}
Experimental evidence for such broadening has been reported for many
superconductors, e.g., Y(Ni$_{0.8}$Pt$_{0.2}$)$_2$B$_2$C, 
\cite{OhishiYEAR03} CeRu$_2$, \cite{KadonoYEAR01} and 
YBa$_2$Cu$_3$O$_{6.95}$, \cite{RisemanYEAR95}. 
On the other hand, to explain the observed
narrowing of the linewidth in some layered high $\kappa$
superconductors, a model in which point vortices randomly displaced
within layers with no alignment or correlation between the layers was
developed.\cite{BrandtYEAR91} In some cases, this could be
topologically similar to the random pinning of segments of flexible
vortex lines. If so, then this effect would be most dominant at
relatively low applied fields, where the interaction between vortices is
weakest. The fact that we observe the same linewidth in the aged sample
at both $H_0$ = 60 mT and 0.3 T seems to argue against this scenario. 
Furthermore, this model for extreme linewidth narrowing was developed
for a highly anisotropic (layered) superconductor. The effective mass anisotropy 
in PuCoGa$_5$ can be estimated from the values\cite{ColineauYEAR06} of the slopes of the upper critical fields 
for $H \parallel c$- and $a$-axes. \cite{GrafYEAR1993} 
These data yield only a small mass anisotropy of about 60-70\%, and thus PuCoGa$_5$ is not a candidate for such a model. We, therefore, conclude that the most likely cause of the sharply reduced linewidth in aged PuCoGa$_5$ is impurity scattering. Note that since any FLL distortion is, therefore, assumed to broaden the linewidth, the reduction in superfluid density deduced for the aged sample is therefore a lower limit; it could be greater.

\section{Modeling the data}
\subsection{Dirty $d$-wave model}
In a superconductor whose electrons are paired in an $L = 0$ ($s$ wave)
orbital angular momentum state, $\rho_{\rm s}$ and
$\Delta\lambda \equiv \lambda(T)-\lambda(0)$ are relatively
temperature-independent below about $T/T_{\rm c} = 0.3$, reflecting
exponentially-activated quasiparticle excitations over a superconducting
gap which is non-zero over the entire Fermi surface.\cite{Tinkham} This
is clearly not observed in PuCoGa$_5$, as seen in Figs. \ref{fresh} and
\ref{aged}, where the behavior of $\rho_{\rm s}$ and $\Delta\lambda$ in both
the fresh and aged samples is approximately linear at low temperatures. 

A pairing state with relative angular momentum $L = 2$ ($d$ wave) has
been found to produce $\rho_{\rm s} \propto T$ at low temperatures in
the clean high-temperature copper-oxide
superconductors.\cite{SonierYEAR00} We have, therefore, compared the
data in Figs. \ref{fresh} and \ref{aged} to model calculations for the
superfluid density $\rho_{\rm s}$ in a $d$-wave superconductor. The
calculations follow from the standard response formula
\cite{KimYEAR94,ChoiYEAR89} for quasi-two-dimensional quasiparticles with a
cylindrical Fermi surface that relates the superfluid density with the
penetration depth,
$\rho_{\rm s}(T) \propto 1/\lambda(T)^2$, 
\begin{eqnarray}
\frac{1}{\lambda(T)^2} = \frac{\omega_p^2}{c^2}
\, k_B T \sum_{m=-\infty}^{+\infty}
 \int_0^{2\pi}\!\!{d\phi}
 \frac{ \cos^2 \phi \, \Delta(\phi;T)^2 }{ [ \tilde\varepsilon_m^2 + \Delta(\phi;T)^2 ]^{3/2} } ,
\label{density}
\end{eqnarray}
where $\omega_p$ is the Drude plasma frequency, $c$ is the speed of
light, and $\tilde\varepsilon_m$ is the impurity renormalized Matsubara
frequency $\varepsilon_m=(2m+1)\pi k_BT$. The impurity self-energy is
calculated within the framework of the $T$-matrix
approximation.\cite{XuYEAR95} For the clean case this expression
simplifies to $1/\lambda(0) = \omega_p/c$. Here $\Delta(\phi;T)$ is the
phenomenological gap function given by $\Delta(\phi;T) =
\Delta_0 \tanh(b\sqrt{T_c/T-1}) \cos 2\phi$,
and $\Delta_0$ and $b$ are phenomenological parameters.\cite{BangYEAR04,HirschfeldYEAR88}
We used $b=1.65$, as for the case of weak-coupling $d$-wave
superconductivity, though our results are not very sensitive to this
parameter. The relevant scales are set by the magnitude $\Delta_0$ at
$T=0$ and the $T$-dependence near $T_c$: 
$\Delta(T) \sim b\Delta_0 \sqrt{1-T/T_c}$.
The model calculations
include the effects of strong impurity scattering to
account for the radioactive decay of the Pu atom. Although PuCoGa$_5$ is not strictly 
a two-dimensional (2D) superconductor, electronic structure calculations predict
several cylindrical Fermi surfaces.\cite{TanakaYEAR04,HottaYEAR03,OpahleYEAR03}
Thus, 2D model calculations for the temperature dependence
of the in-plane $\lambda(T)$ should yield qualitatively reasonable comparisons with
the data. In what follows, we refer to this model as the `dirty $d$-wave' model.

\begin{table*}
\caption{Parameters derived from comparing $\mu$SR rates $\sigma_{\rm v}(T)$ in
 25 day-old (fresh) and 400 day-old (aged) PuCoGa$_5$ to a model of
 dirty $d$-wave superconductivity with gap $\Delta_0$, penetration depth
 $\lambda(0)$ (at $T = 0$), impurity scattering rate $\Gamma$, and the
 transition temperature of the nominally pure sample $T_{c0}=18.5$
 K. Note that within AG pair-breaking theory the suppressed $T_c=15.0$ K corresponds
 to $\Gamma=0.074 \pi k_BT_{c0}$, which is more than seven times bigger
 than the $\Gamma$ needed to fit the $T$ dependence of $\rho_s$.}
\begin{ruledtabular}
\begin{tabular}{cccccc}
Samples & $T_c$ (K) & $\sigma_{\rm v}(0)$ ($\mu$s$^{-1}$) & $\lambda(0)$ (nm) & $n=\Delta_0/k_BT_{c0}$ & $\Gamma/\pi k_BT_{c0}$ \\ \hline
Fresh  & 18.25(10)& 1.52(5)    & 265(5)   & 3.0(1)     & 0.005       \\ 
Aged  & 15.0(1)	& 0.43(1)    & 498(10)   & 2.4(1)     & 0.010       \\
\end{tabular}
\end{ruledtabular}
\end{table*}
The data for the fresh sample are compared to the model
calculations in Fig. \ref{fresh}, where $\sigma_{\rm
v}(0)$ and $T_c$ were used as adjustable parameters. Curves for a range
of gap parameters $\Delta_0$ are drawn as solid lines to show the
sensitivity of the data to the model parameters. Good overall agreement with the data is obtained with this model using a scattering rate $\Gamma = 0.005\pi k_BT_{c0}$, a value consistent with the reduction in $T_c$ according to conventional AG pair-breaking theory (see below). The measured penetration
depths and derived parameters are shown in Table I. 
It is worth noting that the penetration depth of our fresh sample 
($\lambda(0)=265(5)$ nm) is in good agreement with NMR measurements ($\lambda(0)\approx 250$ nm).\cite{CurroYEAR06}
The parameter $n$ in the relation $\Delta_0 =
nk_BT_{c0}$ is an indirect measure of the coupling strength of the
pairing. For a clean weak-coupling $d$-wave state $n = 2.14$. Our best fit to the fresh sample data yields $n=3.0(1)$, indicating an enhanced coupling strength. 

The temperature dependence of $\rho_{\rm s}(T)$, or $\Delta\lambda(T)$,
exhibits an essentially linear behavior at low temperatures. There are at least two phenomena which might be expected to
destroy this linearity in a $d$-wave superconductor. The first is
non-local dynamics in clean
superconductors,\cite{KosztinYEAR97,AminYEAR00} which causes a $T^3$
temperature dependence in $\rho_{\rm s}$ below a temperature 
$k_BT^* \approx (\xi/\lambda(0)) \Delta_0$. We estimate $T^* \approx
(2\, {\rm nm}/265\, {\rm nm}) \cdot 3 \cdot 18.5\, {\rm K} =
0.4$~K. Thus, we do not anticipate non-local effects to be significant
in our measurements. 

Strong impurity scattering can induce a $T^2$ dependence in
$\Delta\lambda(T)$ at low temperatures in a $d$-wave
superconductor.\cite{GrossYEAR86,ChoiYEAR88,ArbergYEAR93,HirschfeldYEAR93}
Conventional AG pair-breaking theory, originally developed for magnetic impurities in an s-wave
superconductor,\cite{AGYEAR61} has also been found to be applicable to
non-magnetic impurities in a $d$-wave superconductor.\cite{Tinkham} The
reduction in $T_c$ for $\Delta T_c/T_c \ll 1$ in the presence of
impurity scattering is given by $\Delta T_c = \pi\Gamma/4$. For our
fresh sample $\Delta T_c \cong 0.25$ K, yielding $\Gamma \cong 0.005
\pi k_BT_{c0}$, the scattering rate used to model the data in
Fig.~ \ref{fresh}. The cross-over temperature to $T^2$ behavior for
unitary (strong) scattering has been estimated\cite{HirschfeldYEAR93}
(independently from the model\cite{BangYEAR04,HirschfeldYEAR88} used in
Figs. \ref{fresh} and \ref{aged}) to be $T_{\rm cr}\approx
0.83\sqrt{\Gamma \Delta_0}$, where $\Delta_0$ is the maximum gap
amplitude. Taking $\Delta_0 = 3k_BT_{c0}$ for the fresh sample yields
$T_{\rm cr} \cong 3.3$ K. The crossover from $T$-linear to $T^2$
behavior is expected to be smooth, so that $\Delta\lambda(T) \approx
bT^2/(T + T_{\rm cr})$.\cite{HirschfeldYEAR93} Therefore, a $T^2$
behavior is only anticipated for temperatures significantly below
$T_{\rm cr}$; practically, this works out to be $T \leq T_{\rm cr}/3$,
as observed by $\mu$SR\cite{PanagopoulosYEAR96} for Zn doping in
YBa$_2$Cu$_3$O$_{7-\delta}$. Thus, using either our model or this
simpler analytic estimate, we do not expect to see evidence for strong $T^2$
behavior as long as $T \agt 3$ K in fresh PuCoGa$_5$, and we do not. 

The same model has been used to fit the aged sample data, shown in Fig. \ref{aged}, where again a range of gap parameters is displayed. The best fits are obtained for $n=2.4$ and $\Gamma = 0.01\pi k_BT_{c0}$. For comparison, the dotted
curve in Fig.~ \ref{aged} shows the best fit to the fresh sample data
($\Delta_0 = 3k_BT_{c0}$, $\Gamma = 0.005\pi k_BT_{c0}$), indicating
either that the coupling strength has been reduced with aging, or 
that pair-breaking has noticeably reduced the gap value by
about 20\% after 400 days of aging. 
The dashed curve in Fig.~ \ref{aged} is a calculation for a dirty $d$-wave superconductor with $\Delta_0 = 2.4k_BT_{c0}$ and $\Gamma = 0.074\pi k_BT_{c0}$ in
the strong scattering limit. This value of $\Gamma$ is required by conventional AG pair-breaking theory to reproduce the reduction of $T_c$ in the aged sample, but it clearly leads to poor agreement with the 
temperature dependence of the superfluid density. Furthermore, as we now discuss, this 
 model does not account for the strong reduction in $\rho_s$ in the aged
sample.


\subsection{Short coherence length model}

In our discussion above, we concluded that the most likely cause of the sharply
reduced linewidth in aged PuCoGa$_5$ is impurity scattering. This
situation is not unprecedented. In comparing doped and radiation-damaged
YBa$_2$Cu$_3$O$_{7-\delta}$ (YBCO) superconductors to the conventional
AG pair-breaking theory, it was found in many cases that $T_c$ was remarkably robust, even though
$\rho_{\rm s}$ was easily
suppressed.\cite{UlmYEAR95,BasovYEAR94,MoffatYEAR97,NachumiYEAR96,BernhardYEAR96} In Ni-doped and
He-irradiated YBCO materials, for example, a suppression of $T_c$ by
about 20\% was accompanied by a suppression of $\rho_{\rm s}$ by about
70\%. This large $T_c/\rho_{\rm s}$ ratio is in contradiction to
conventional `dirty $d$-wave' theory, and has been addressed
theoretically by accounting for the suppression of the order parameter
around the vicinity of the
defects.\cite{FranzYEAR97,ZhitomirskyYEAR98,HettlerYEAR99} 
The conventional treatment assumes that the spatial variation of the
superconducting order parameter (of order $\xi$) is large compared to
the average distance $\bar{l}$ between the
scattering centers, but large compared to the lattice parameter $a_0$, so that the effective order parameter is a spatial
average, instead of having a large suppression only near the
impurities. 
As expressed by Franz {\it et al.}\cite{FranzYEAR97}, the effect of a spatially inhomogeneous gap is
enhanced by a short coherence length relative to the lattice parameter,
$\xi/a_0\approx$ 2 - 5.\cite{FranzYEAR97} As in YBCO, the coherence
length in PuCoGa$_5$ is relatively small ($\sim 2$ nm in both materials)
compared to many superconductors where conventional AG pair-breaking theory
applies. Although we do not know the actual spacing $\bar{l}$ between
defects after self-annealing, we can estimate the mean distance $d$
between Pu atoms which have decayed randomly after 400 days as a crude
measure of the spacing between damage cascades. We find $d \cong 20$ nm,
which is $\approx 10 \xi$. Thus, a spatially inhomogeneous gap model
with a short coherence length, as considered by Franz and
coworkers,\cite{FranzYEAR97} may explain our observations.

In this regard, we note that both the suppression of $T_c$ with aging
and the value of $\xi$ are about twice as large in PuRhGa$_5$ compared
to PuCoGa$_5$.\cite{SakaiYEAR05,BangYEAR06a,BangYEAR06b} 
Very recently, an enhanced suppression of the superfluid density with
chemical doping has also been reported for the related compound
CeCoIn$_{5-x}$Sn$_x$.\cite{BauerYEAR06}

In order to test our hypothesis of short-coherence-length superconductivity for
PuCoGa$_5$, we calculated the superfluid density by solving the Bogoliubov-de Gennes (BdG) equations of a 2D $d$-wave superconductor:
\begin{equation}
\sum_{j} \left(
\begin{array}{cc}
\mathcal{H}_{ij} & \Delta_{ij} \\
\Delta^{*}_{ij} & - H_{ij}^{*} 
\end{array}
\right) 
\left(
\begin{array}{c} u_{j}^{n} \\ v_{j}^{n} \end{array}
\right) 
=E_{n} \left( 
\begin{array}{c} u_{i}^{n} \\ v_{i}^{n} \end{array}
\right) \;.
\end{equation}
Here $(u_{i}^{n},v_{i}^{n})^{T}$ are the eigenfunctions at site $i$ corresponding to the 
quasiparticle excitation energy $E_{n}$. 
The normal-state single-particle lattice Hamiltonian is 
\begin{equation}
\mathcal{H}_{ij} = -t \delta_{i+\delta,j} + (U_{\text{imp}}- \mu) \delta_{ij} \;,
\end{equation}
where $t$ is the hopping integral between a specific lattice site and 
its four nearest neighbors as denoted by 
$\delta=(\pm 1, 0)$ and $(0, \pm 1)$, 
$\mu$ is the chemical potential, and $U_{\text{imp}}$ is the impurity potential 
modeling the on-site disorder. 
The self-consistency equation for the gap function of a 
$d$-wave superconductor on a lattice is given by 
\begin{equation}
\Delta_{i,j=i+\delta}= \frac{V}{2} \sum_{n} [ u_{i}^{n} v_{j}^{n*} + u_{j}^{n}v_{i}^{n*}]\tanh(E_{n}/2k_{B}T)\;,
\end{equation}
with $V$ being the pairing strength.
Note that the quasiparticle energy is measured with respect to the chemical potential. 
We follow an iterative procedure to solve self-consistently the BdG equations by exact 
diagonalization on a $20 \times 20$ lattice, and, for comparison, on a $24 \times 24$ lattice.
Using a suitable guess for an initial $\Delta_{ij}$, a new one is then 
calculated, and the process is iterated until the desired convergence is achieved. 

\begin{figure}[tp]
\rotatebox[origin=c]{0}{\includegraphics[width=0.8\columnwidth]{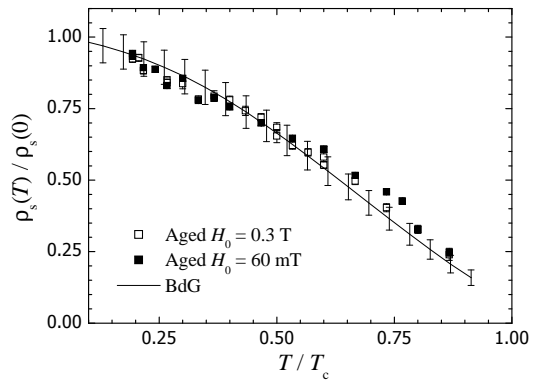}}
\caption{Temperature dependence of the normalized superfluid density 
 $\rho_s(T)/\rho_s(0)=\lambda(0)^2/\lambda(T)^2$ in PuCoGa$_5$. 
 The solid squares are for $H_0 = 60$ mT and the open squares for $H_0 = 0.30$ T 
 after 400 days, using $\sigma_{\rm v}(0) = 0.45(1)$ $\mu$s$^{-1}$
corresponding to $\lambda(0) = 487(10)$ nm. The solid line is the short-coherence-length BdG superfluid density 
 (for $\rho_s(0;n_{\rm imp}=6\%)/\rho_s(0;n_{\rm imp}=0\%) \approx 0.35$) in the strong scattering limit described by 
 $\Delta_0 = 0.187 t$, $T_c = 0.115 t$, 
 and impurity concentration $n_{\text{imp}}=6\%$. 
For the pure case, $n_{\text{imp}}=0\%$, one finds
 $\Delta_0 = 0.314 t$ with $T_{c0}=0.143 t$.
 The statistical error bars are due to averaging over 20 random impurity configurations at 
 fixed concentration.}
\label{aged_BdG}
\end{figure}

The linear response calculation of the superfluid density $\rho_s$ for the 
lattice model follows the approach 
described by Scalapino et al.~\cite{ScalapinoYEAR92} and was applied
to 2D $d$-wave \cite{XiangYEAR95,FranzYEAR97} 
and $s$-wave~\cite{GhosalYEAR98,GhosalYEAR01} superconductors.
A detailed technical description of the response calculation will 
be given elsewhere.\cite{ZhuYEAR07}
Our calculations of the transition 
temperature $T_c$, the lattice-averaged zero-temperature order parameter 
$\Delta_0 = \langle \Delta_{ij}(0) \rangle$
and superfluid density $\rho_s(0)$ are in good agreement with those of Franz et al.\cite{FranzYEAR97}
for the same set of parameters as used by them ($V=1.13 t, U_{\text{imp}}=100 t, \mu=-0.36 t$).
Because of finite size effects of our lattices, we averaged our calculations over 20 random
impurity configurations at fixed concentration, as indicated by the statistical
error bars in Fig.~\ref{aged_BdG}.

In Fig.~\ref{aged_BdG} we compare the temperature dependence of $\rho_s(T)$
for the aged sample and the BdG response calculation in the limit of strong impurity
scattering. The impurity concentration $n_{\text{imp}} = 6\%$ was chosen to reproduce
the observed suppression of $T_c$ and $\rho_s(0)$ of the aged vs.\ fresh sample. The relevant 
parameters are given in the caption to Fig.~\ref{aged_BdG}. The BCS coherence length 
is given by $\xi_0 \sim \hbar v_F/(\pi\Delta_0)$, where the Fermi velocity is given approximately by 
$v_F \sim a_0 t/\hbar$ with $\Delta_0 = 0.187 t$, so that $\xi_0 \sim 2a_0$.
The excellent agreement between the short-coherence-length BdG calculation and the
measured superconducting properties of the aged sample, and, at the same time, 
the failure of the dirty $d$-wave calculation within AG pair-breaking theory to describe the data, 
indicates that the uniform, dilute-impurity theory by Abrikosov and Gor'kov is 
not applicable to PuCoGa$_5$, where superconductivity is seemingly
not uniformly suppressed. Specifically, both the BdG and AG dirty $d$-wave models require $k_f \bar{l} \gg 1$, where $k_f$ is Fermi wave vector. However the former case requires strong impurity scattering with on-site suppression of $\Delta_0$ and $k_f \xi \sim 1$, while the AG pair-breaking theory requires a uniformly suppressed $\Delta_0$ with $\xi/\bar{l} \alt$ 1.

\section{Discussion and Summary}

We now summarize our principal results. The radioactive decay of $^{239}$Pu allows one to study the effects of pair-breaking defects on the properties of the superconductivity in PuCoGa$_5$. The low-temperature,
quasi-linear temperature dependences of the superfluid density and penetration depth in both 
fresh (25 day-old) and aged (400 day-old) PuCoGa$_5$ are consistent with a line of nodes in
a $d$-wave order parameter. 

The fresh sample is almost defect free, as evidenced by the small reduction in $T_c$, e.g., $\Delta T_c/T_{c0} \approx$ 1.4\%. We have, therefore, compared our data in the fresh sample to a dirty $d$-wave model for 
a strong-coupling superconductor in the presence of strong impurity scattering. We find that the zero-temperature penetration depth and superconducting gap are $\lambda(0) = 265(5)$ nm and $\Delta_0= nk_BT_{c0}$ with $n =
3.0(1)$, respectively. The latter is enhanced compared to the weak-coupling $d$-wave case, where $n =
2.14$. The model scattering rate used was consistent with the slight reduction in $T_c$ for the fresh sample, according to conventional
AG pair-breaking theory. Our results on the fresh sample are 
consistent with recent NQR and NMR experiments in fresh
PuCoGa$_5$.\cite{CurroYEAR05,CurroYEAR06} 

In the case of the aged sample, we argued that the presence of strong pinning in PuCoGa$_5$ allows an accurate measure of the temperature dependence of $\lambda(T)$, but that distortion of the FLL (which broadens the linewidth) means that the measured magnitude of $\lambda(0)$ is only a lower limit. The dirty $d$-wave
model can also fit the temperature dependence of the superfluid density $\rho_s(T)$ (or $\lambda(T)$) with $\Delta_0$ reduced by about 20\%, but with
a scattering rate nearly an order of magnitude smaller than predicted by conventional AG pair-breaking theory. 
Furthermore, the dirty $d$-wave model is unable to account for the reduction of at least 70\% in $\rho_s(0)$. This is attributed to the fact that PuCoGa$_5$ possesses a relatively short coherence length, and, therefore, the conventional AG pair-breaking theory, in which the order parameter is spatially averaged over impurity sites, is inappropriate. This was pointed out by Franz {\em et al.}\cite{FranzYEAR97} for damaged or doped YBCO superconductors, but in that work a comparison with the data was made only for zero temperature. 

Accordingly, we have modeled the full temperature dependence of $\rho_s(T)$ in aged PuCoGa$_5$ by solving the BdG equations for a short-coherence-length, weak-coupling $d$-wave superconductor in the strong scattering limit. (For small scattering rates the short-coherence-length model agrees fairly well with the dirty $d$-wave model.\cite{FranzYEAR97}) The BdG model is able to reproduce quite well the temperature dependence and magnitude of $\rho_s$ for a nominal 6\% impurity concentration, chosen to reproduce the magnitude of $\rho_s(0)$. This impurity concentration is reasonable, though we attach no particular importance to its exact value. Like the dirty $d$-wave model, the BdG model is consistent with a reduction of $\Delta_0$ in the aged material, compared to the fresh sample. Both the dirty $d$-wave and BdG models are consistent with $\lambda(0) = 498(10)$ nm in aged PuCoGa$_5$, where we remind the reader that because of FLL distortion this is a lower limit. 

Computational feasibility dictates that the BdG lattice model calculations are for a two-dimensional system. Therefore, the derived gap parameters are only semi-quantitative. (The values of $\lambda(0)$ are accurately determined from Eq. (\ref{lambda}), however, and our quoted values are derived from this expression.) Nevertheless, we do not expect strong deviations between 2D and 3D models for the in-plane penetration depth, and, therefore, believe that a 2D model should provide a good qualitative description of the essential physics in this interesting superconductor. \cite{GrafYEAR1996} 

Our data and analysis
in aged PuCoGa$_5$ suggest that although small parts of the
sample are significantly disordered, and consequently the superfluid density 
is strongly suppressed in these regions, 
superconductivity remains remarkably resilient.

\section*{Acknowledgments}
This work was partially supported by a Grant-in-Aid for Scientific Research (No. 18027014), the Ministry of Education, Culture, Sports, Science and Technology, Japan. Work at LANL (contract No.\ DE-AC52-06NA25396) and LLNL (contract No.\
W-7405-Eng-48) was performed under the U.S.\ Department of Energy. Work
at Riverside was supported by the U.S.\ NSF, Grant DMR-0422674.
We thank the staff of TRIUMF and acknowledge helpful discussions with I. Affleck, M. Franz, 
G. Lander, P. M. Oppeneer, J. E. Sonier, J. D. Thompson, and F. Wastin.


\end{document}